\documentclass{ieeeaccess}
\usepackage{cite}
\usepackage{amsmath,amssymb,amsfonts}
\usepackage{algorithmic}
\usepackage{graphicx}
\usepackage{textcomp}
\usepackage{subfigure}

\def\BibTeX{{\rm B\kern-.05em{\sc i\kern-.025em b}\kern-.08em
    T\kern-.1667em\lower.7ex\hbox{E}\kern-.125emX}}
\begin{document}
\history{Date of publication xxxx 00, 0000, date of current version xxxx 00, 0000.}
\doi{}

\title{Towards Fault Diagnosis in Induction Motor using Fractional Fourier Transform}
\author{\uppercase{USMAN ALI}}

\begin{abstract}
A method for determining the current signature faults using Fractional Fourier Transform (FrFT) has been developed. The method has been applied to the real-time steady-state current of the inverter-fed high power induction motor for fault determination. The method incorporates calculating the relative norm error to find the threshold value between healthy and unhealthy induction motor at different operating frequencies. The experimental results demonstrate that the total harmonics distortion of unhealthy motor is much larger than the healthy motor, and the threshold relative norm error value of different healthy induction motors is less than 0.3, and the threshold relative norm error value of unhealthy induction motor is greater than 0.5. The developed method can function as a simple operator-assisted tool for determining induction motor faults in real time.          
\end{abstract}


\titlepgskip=-15pt

\maketitle

\section{Introduction}
Three-phase induction motors are widely used in industrial processes due to their excellent performance, low maintenance cost, ruggedness, robust structure, and ability to function under harsh environments \cite{r1,u1}. Therefore, induction motors are estimated to account for $\sim$85\% share of the motors employed in the industry \cite{r2}. Recently, variable frequency drives (VFDs) are increasingly being employed to control and regulate the operations of induction motors, and depending upon the load requirement, the power consumption of a motor can be reduced by 75\% by operating it at the half-speed \cite{r3}. 

The unwanted electrical and mechanical effects disturb the performance of an induction motor, consequently affecting the production line in time and monetary losses \cite{r4,r5,u2}. The common faults reported in induction motors include bearing related faults, stator faults, and rotor faults; whereas the bearings faults have been reported to constitute a larger proportion of $\sim$40\% of the overall faults \cite{Singh2003, Albrecht1986}. Therefore, the identification and classification of faults in induction motors is an attractive research area that has been investigated using various conditioning monitoring techniques. These techniques assist in predicting the motor failures, optimize motor maintenance, reduce maintenance costs, reduce downtime, and improve the overall reliability of the system \cite{r6, r7, r8, r9}. Various approaches can be used for identification and classification of faults in the induction motor, including thermal analysis, chemical analysis, acoustic analysis, torque analysis, induced voltage analysis, partial discharge analysis, vibration analysis, and motor current signature analysis (MCSA) \cite{r16, r18, r19, r20,u3}. Secondly, various signal processing techniques have been reported in the literature for the induction motor condition monitoring, including fast fourier transform (FFT), short-time fourier transform (STFT), wavelet transform (WT), hilbert transform (HT), Wigner–ville distribution (WVD), adaptive slope transform (AST), chirplet transform (CT), and FrFT \cite{r9, r11, r12, r13, r14, r15, u4}. 

Motor current signature analysis (MCSA) is widely used for fault detection and classification in induction motors. Different types of faults such as broken rotor bar fault, air gap eccentricity, stator fault, abnormal connection of stator winding, bent shaft, bearing fault, and gear-box fault can be determined using MCSA \cite{Yang2016,r21,r22}.

\begin{figure}[ht!]
    \includegraphics[width=\columnwidth]{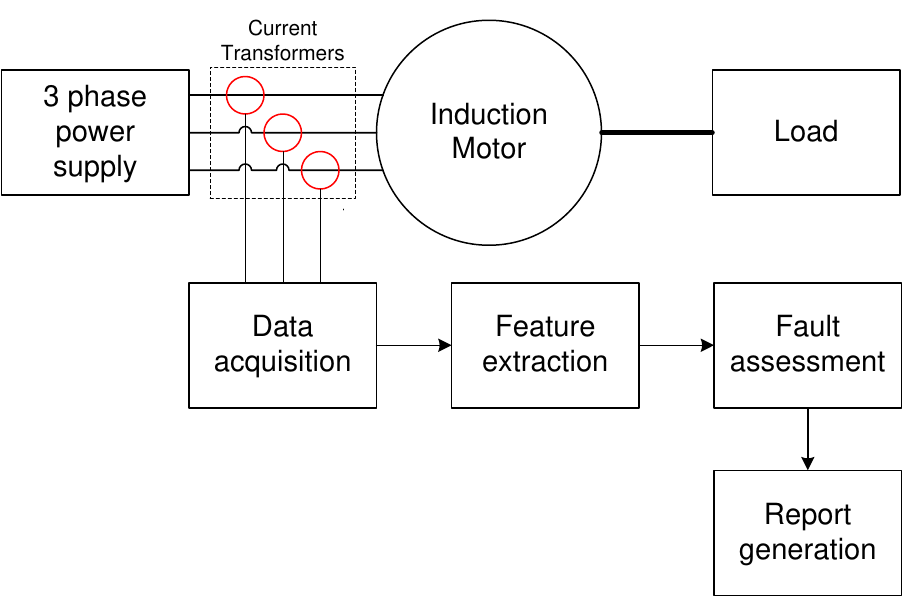}
    \caption{The flow diagram of MCSA technique. The stator current is the fundamental component for fault detection and diagnosis.}
    \label{fig1}
\end{figure}

MCSA technique is divided into three steps, 1) data acquisition, 2) feature extraction, and 3) fault assessment. The flow diagram of MCSA technique is shown in figure \ref{fig1}.

The Stator current of the induction motor is measured using a current transformer (CT) in the data acquisition step. Normally one phase is enough for extracting the required information about the signal, but for more accuracy, we can monitor three-phase current using hall effect or CT. The signal acquired from the phase current is in time-domain, therefore, it is converted to frequency domain data using FFT. In the feature extraction stage, various signal processing and machine learning techniques can be used to extract significant features from the recorded data. FFT, STFT, WT, HT, SVM, and Park's vector approach can be used for signal processing. In the Fault assessment stage, we compare the frequency spectrum between healthy and unhealthy motors for generating accurate reports about motor faults.

In this paper, we present the application of FrFT method for fault identification in induction motors. The recorded motor current signal data has been used to calculate the relative norm error against varying rotation angle of FrFT for the healthy and faulty induction motors. This has resulted in obtaining a threshold relative norm error value which can be used to identify a faulty induction motor. The developed method presents a simple alternate to the online and offline condition monitoring techniques.

The rest of the paper is organized as follows; section \ref{faults} presents various types of faults that may occur in induction motors and mathematical relations between FFT and FrFT, section \ref{Proposed} presents our proposed approach, section \ref{results} presents the experimental results and discussion, and finally, section \ref{concl} concludes our work.

\section{Faults in Induction Motor}
\label{faults}

The induction motors faults can be generally divided in to following categories:

\begin{itemize}
    \item Broken rotor bar fault
    \item Bearing fault
    \item Air gap eccentricity fault
    \item Stator winding fault
  \end{itemize}

\subsection{Broken rotor bar fault}
\label{brokenrotor}
Thermal and mechanical stresses cause the broken rotor bar fault in induction motors. Heat and power losses occur in the induction motor when few bars are cracked. If left unattended, broken rotor bar fault can easily damage the motor winding\cite{r23}. This type of fault can be detected in early stages by analyzing the frequency spectrum of the stator current where different sidebands are produced around the fundamental frequency component of the supply. Sideband frequency components can be calculated using equation \ref{equation1} \cite{Guajardo2018}.

\begin{equation}
    f_\text{brb}=f_\text{f}(1 \pm 2ks)
    \label{equation1}
\end{equation}

where $f_\text{f}$ is the fundamental frequency, $s$ is slip, and $k$ is an integer.
 
\begin{figure}[ht!]
    \includegraphics[width=\columnwidth]{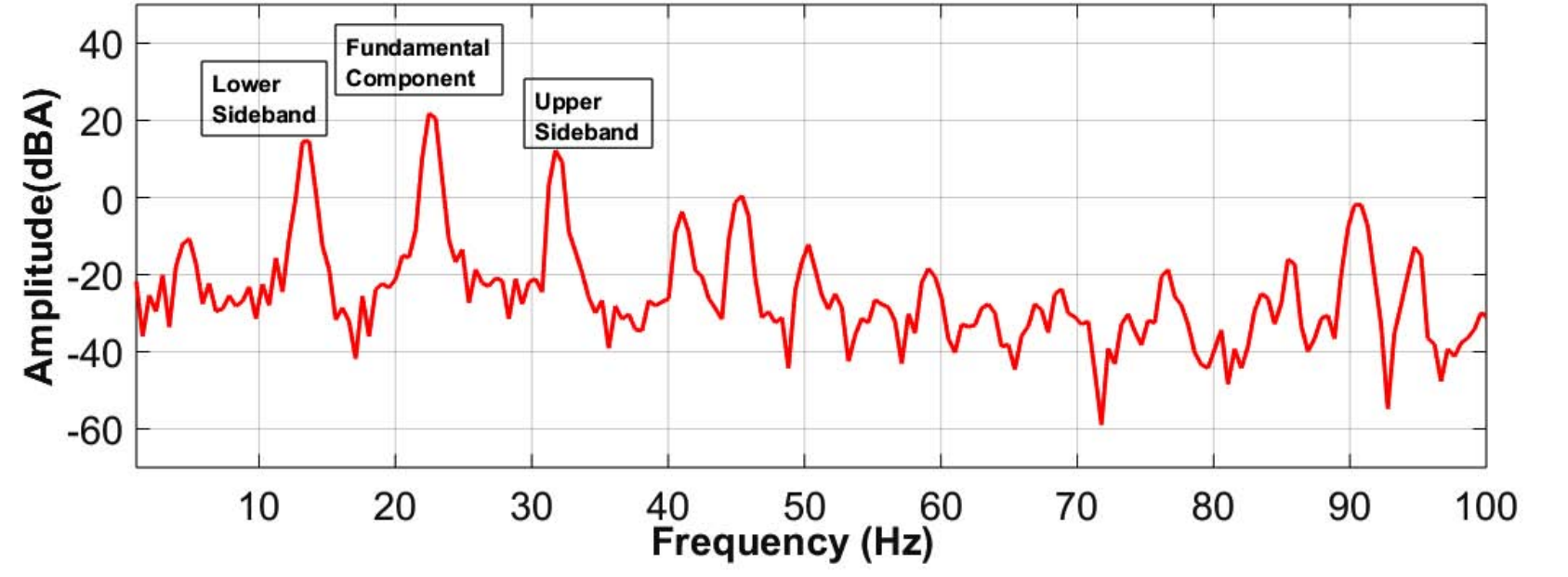}
    \caption{FFT spectrum of the induction motor stator current. The operating frequency of the motor is $\sim$22 Hz. The upper and lower sidebands depict the broken rotor bar faults in the motor.}
    \label{fig2}
\end{figure}

FFT spectrum of the induction motor stator current having broken rotor bar fault is shown in figure \ref{fig2}. The signal is recorded from 20 hp motor operating at $\sim$22 Hz with load. The current signal has been acquired using 1:1/2000 CT. 
 
It is observed in literature that if the difference between the amplitude of the fundamental frequency and the sideband frequency components is larger than 50 db then no broken rotor bar fault exists. Secondly, if the difference between the amplitude of the fundamental frequency and the sideband frequency components is between 40-50 db then one rotor bar is cracked. Finally, if the difference is less the 40 db then multiple rotor bars are cracked \cite{Sharma2017}.

\subsection{Air gap eccentricity fault}
Air gap eccentricity fault occurs in the induction motor if the air gap distribution between the rotor and the stator is not uniform \cite{Hong2012}. Two types of eccentricity faults occur in the induction motor, one is static and the other is dynamic. In case of static air gap eccentricity, nominal air gap is fixed, but in case of dynamic air gap eccentricity, the air gap is distributed according to the rotor's rotation because rotor is not located in the center of stator. Frequency components due to abnormal air gap eccentricity can be calculated using equation \ref{equation2} \cite{Hong2012,r24}.

\begin{equation}
    f_\text{ec} = f_\text{f}\left[(R\pm n_\text{d})\frac{(1-s)}{p}\right]+ n_\text{ws}
    \label{equation2}
\end{equation}

where $f_\text{ec}$ is eccentricity frequency, $f_\text{f}$ is fundamental frequency, $R$ is the number of rotor slots, $n_\text{d} = \pm1$, $s$ is slip, $p$ is the number of pole pairs, and $n_\text{ws}$ is an odd integer.

\subsection{Stator fault}

Stator fault commonly occurs in the stator winding due to inter short turns. When the current is flowing in the stator winding then the short turns produce a back magnetomotive force (MMF) which reduces the net MMF. This also produces heat inside the stator winding. The frequency components due to the short turns in stator winding can be calculated using equation \ref{equation3} \cite{Park2016,r25}.

\begin{equation}
    f_\text{st} = f_\text{f}\left[\frac{n(1-s)}{p}\right]\pm k
    \label{equation3}
\end{equation}

where $f_\text{st}$ is the frequency component of shorted turn, $f_\text{f}$ is fundamental frequency, $n$ is integer, $s$ is slip, $p$ is the number of pole pairs, and $k$ is an odd integer.

\subsection{Bearing fault}

Motor bearings are attached with the end rings of the rotor placed inside the stator winding \cite{Benbouzid1999}. Bearing faults commonly occur due to the improper installation of bearing inside the induction motor. Bearing faults can be diagnosed using the abnormal frequency components calculated using following expressions \cite{Singh2017,r26}.

\begin{equation}
    f_\text{0} = 0.4nf_\text{rm},
    f_\text{1} = 0.6nf_\text{rm}
\end{equation}

where $n$ is the number of balls in bearing, $f_\text{rm}$ is the rotor's mechanical frequency, $f_\text{0}$ is the lower frequency, and $f_\text{1}$ is upper frequency.

\subsection{Mathematical relation between FFT AND FrFT}
The fourier transform of function $f(x)$ is defined as follows:

\begin{equation}
    F(\omega)=\int_{-\infty}^{+\infty} f(x) \text{exp}^{-iwx} dx
\end{equation}

Kernel function of FFT is exp$^{-iwt}$, $i$ equals to $\sqrt{-1}$.

The mathematics of FrFt was presented in literature as early as 1929 \cite{r28,r29,r30}. Later on, it was used in signal processing \cite{r31}, quantum mechanics \cite{r32,r33} and optics \cite{r34}. FrFT of a signal $x(t)$ is defined as \cite{r35}.
\begin{equation}
    F^a(\xi)=\int_{-\infty}^{+\infty} K_a(\xi,x) f(x) dx
    \label{equation4}
\end{equation}
where $a$ is the order of FrFT, $\alpha$ is the rotational angle which is equal to $a\pi/2$, $\xi$ corresponds to fractional time, $a$ and $\xi$ $\in$ $\Re$. $K$ is a kernel function which is defined as follows:

\begin{equation}
    K_a(\xi,x)=C_\text{a} \text{exp} \left[-i\pi(2\frac{x\xi}{sin\alpha} -(x^2+\xi^2) cot\alpha)\right]
\end{equation}

$C_a$ is complex envelope which is defined as follows:

\begin{equation}
    C_a=\sqrt{1-i cot\alpha}=\text{exp} \left[-i(\frac{\pi\text{sgn}(sin\alpha)/4-\alpha/2}{\sqrt{|sin\alpha|}})\right]
\end{equation}

FrFT is the generalization of FFT \cite{r36} because, in FrFT, we rotate a time-domain signal into the frequency domain by varying a rotational angle $\alpha$. If $\alpha$ is equal to zero, then there is no rotation occurred in the signal (which means that signal is still in the time domain). However, if $\alpha$ is equal to one, then the output of the FrFT kernel function is the same as the output of the kernel function of FFT (which means that FrFT becomes FFT at the rotational angle equal to one). Additionally, we can extract different features and remove noise from the original signal by tuning the rotational angle $\alpha$ \cite{r38,r39}.   

\section{Proposed Approach} 
\label{Proposed}

\begin{figure*}[ht!]
\centering
    \includegraphics[width=6in]{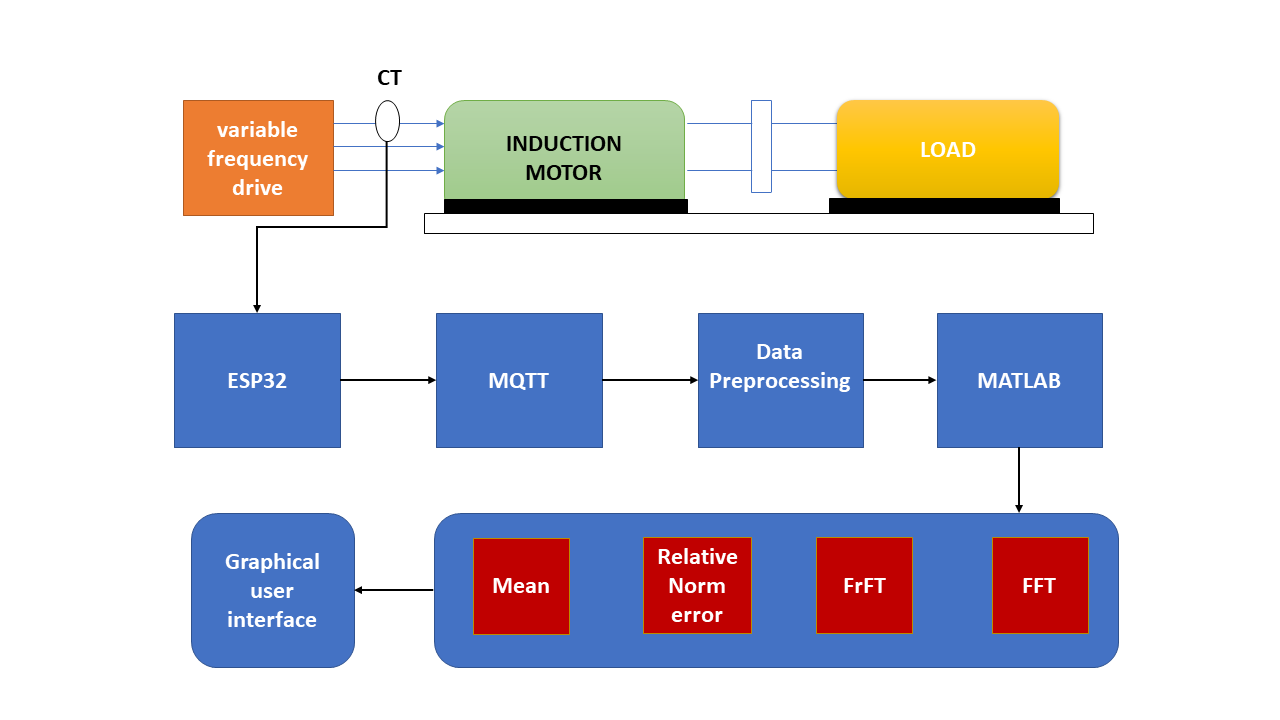}
    \caption{Block diagram of the implemented system. The measured current, temperature, and stage vibration signals are recorded at the micro-controller (ESP32), which then transmits them to the device manager for signal processing and data display.}
    \label{fig3}
\end{figure*}

The block diagram of our implemented system is shown in figure \ref{fig3}. A  variable frequency drive  (VFD)  is used for regulating and control the speed of the induction motor as well as assisting the operator to start the induction motor at different frequencies. Current transformer is used to measure the stator current of the induction motor. The measured stator current signal is recorded at the micro-controller (ESP32), which subsequently transmits wirelessly to the cloud using message queuing telemetry transport (MQTT) communication protocol. Before applying the signal processing techniques, data pre-processing is performed in MATLAB to extract useful features.  In the first instance, the FFT technique is applied to the pre-processed data of the healthy and unhealthy induction motor. FFT converts the time domain signal into the frequency domain spectrum. The study of the harmonic spectrum gives the insights into the health of the induction motor.
Secondly, the FrFT technique is applied to induction motors data of varying the rotational angle $\alpha$. By changing the rotational angle $\alpha$,  the signal can be transformed into an intermediate domain between time and frequency.
After applying the FFT and FrFT, the third step is to measure the relative norm error between various equal length samples of healthy and unhealthy induction motor.  In the end, the mean of the relative norm error is calculated to find the threshold value between healthy and unhealthy induction motor.

For a random variable vector $X$, made up N observations, relative norm error and mean is defined as follows \cite{r40}: 

\begin{equation}
    \text{Mean ($\mu$)} = 
    \frac{1}{N} \sum_{i=1}^{N} X_i
    \label{equation5}
\end{equation}

\begin{equation}
    \text{Relative norm error} = 
    \frac{||x_\text{approx}-x||}{||x||}
    \label{equation6}
\end{equation}

\section{Experimental Results and discussion}
\label{results}
Preliminary functional experiments are performed to test the recorded data of the healthy and unhealthy induction motors.
\begin{figure*}[ht!]
\begin{center}
    
    \subfigure[]{\includegraphics[width=\columnwidth]{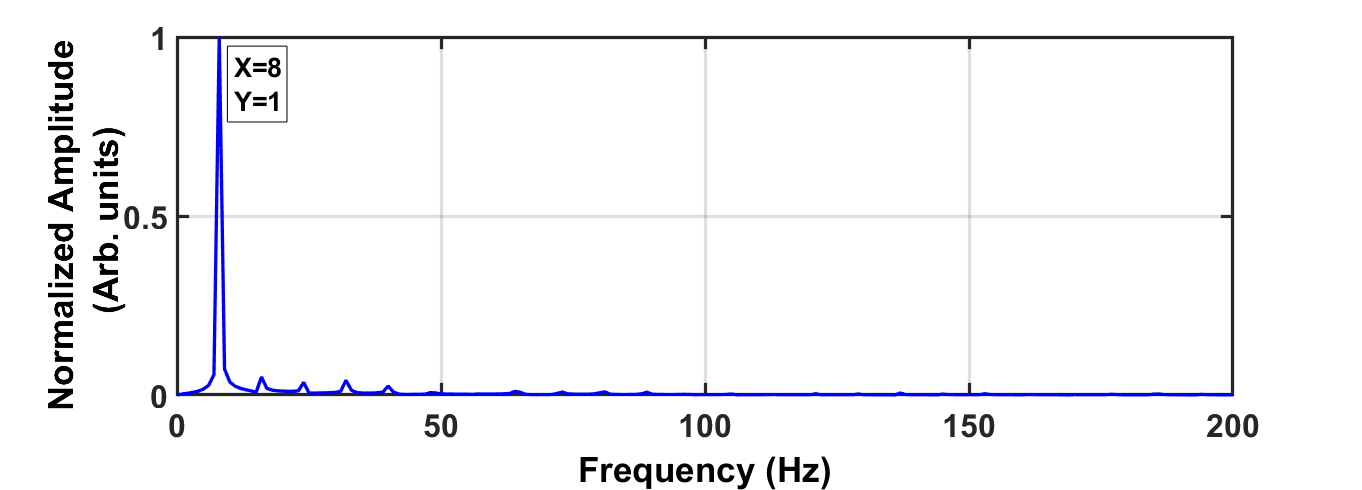}}\subfigure[]{\includegraphics[width=\columnwidth]{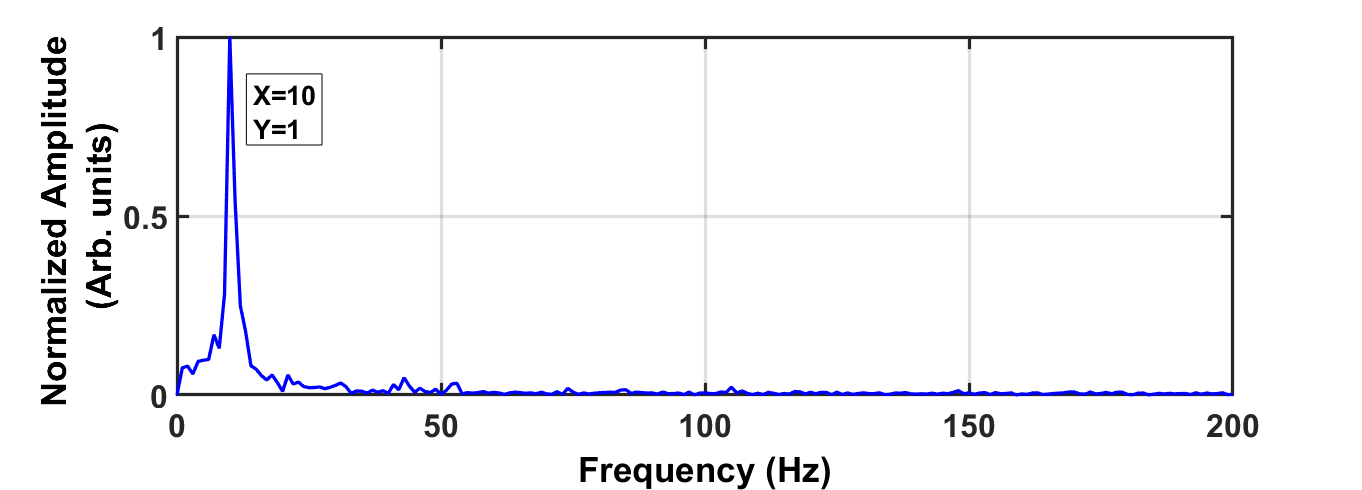}}
    \subfigure[]{\includegraphics[width=\columnwidth]{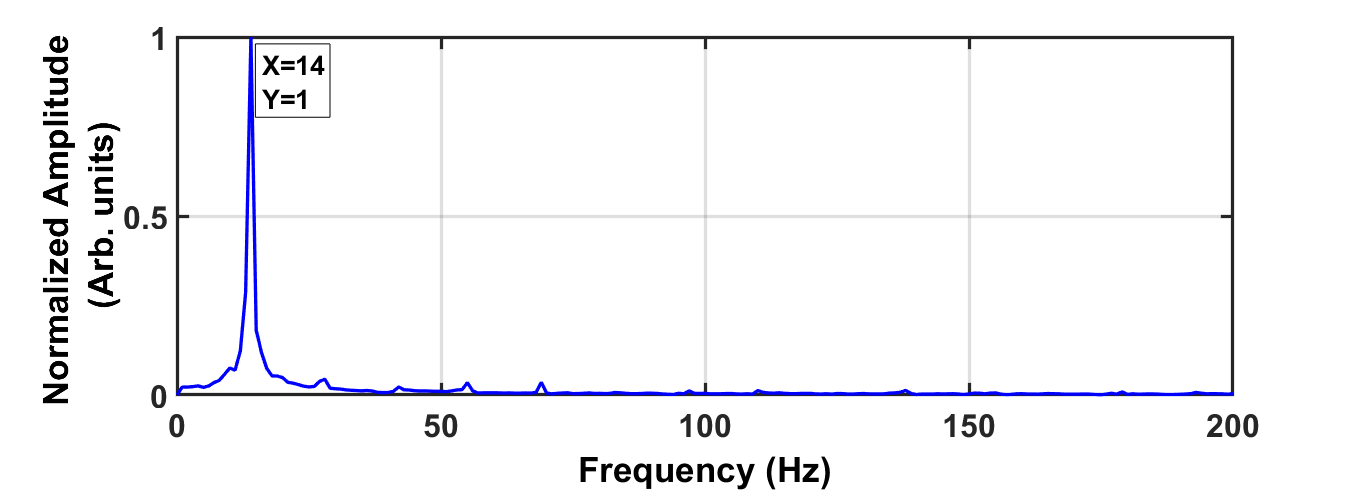}}\subfigure[]{\includegraphics[width=\columnwidth]{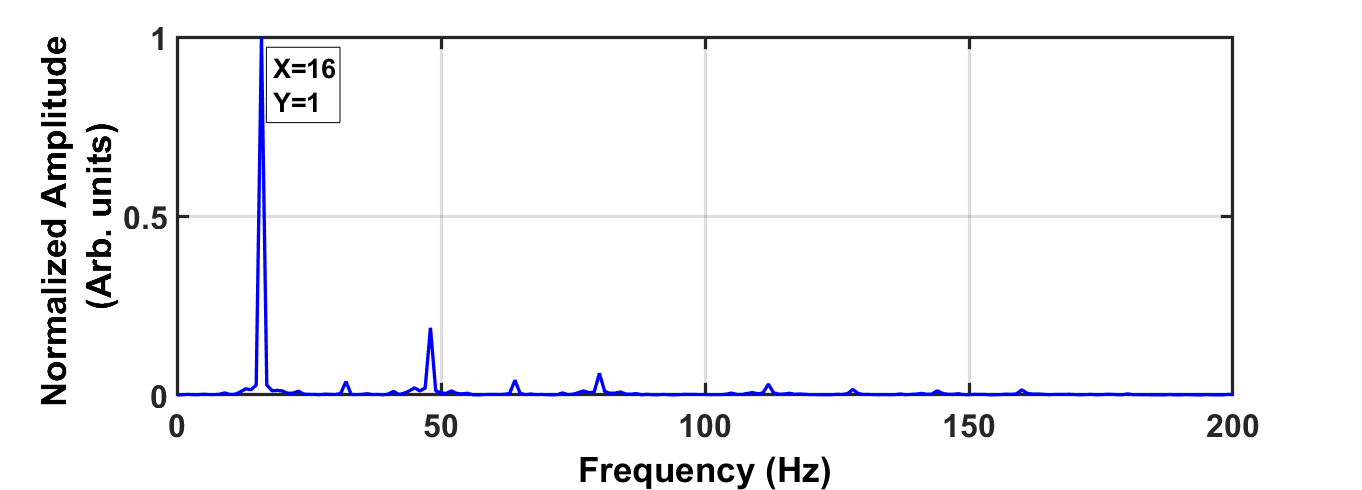}}
    \subfigure[]{\includegraphics[width=\columnwidth]{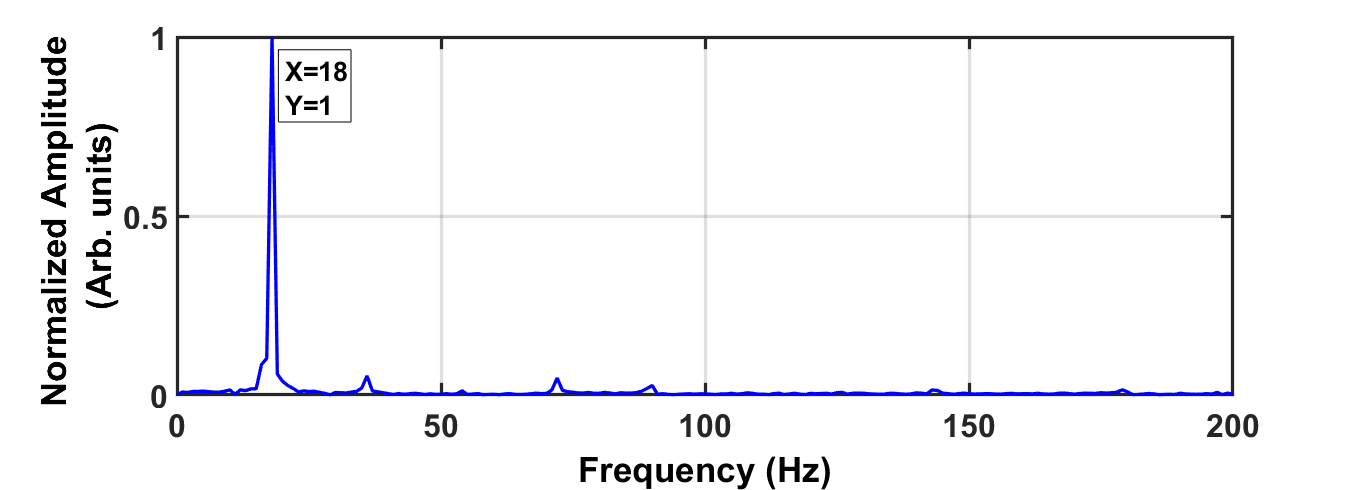}}\subfigure[]{\includegraphics[width=\columnwidth]{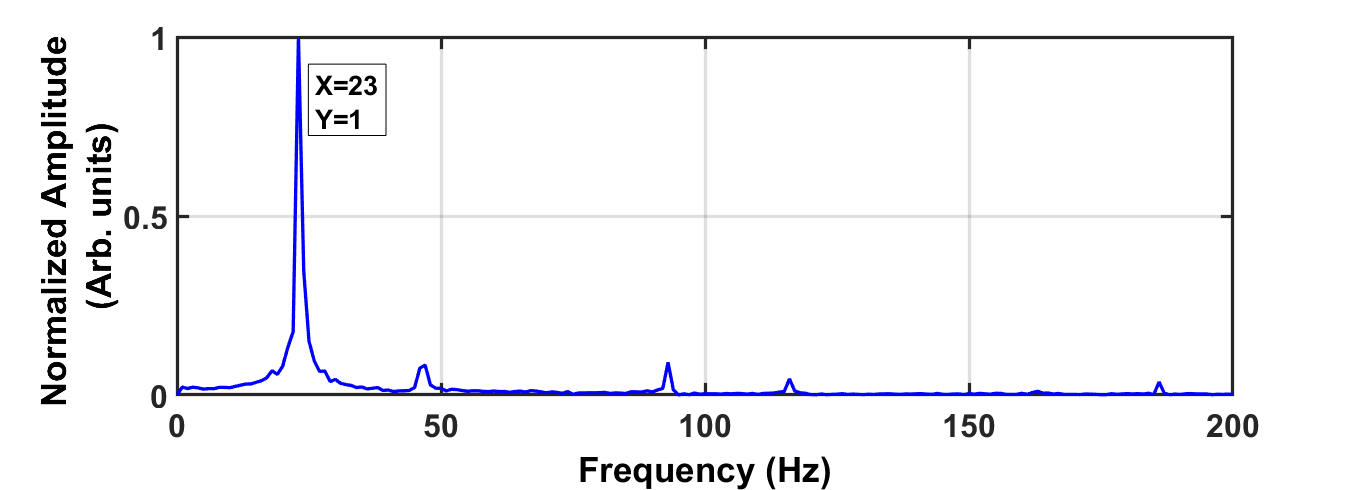}}
    \caption{ Frequency spectrum of healthy 20 hp induction motor operating at (a) 8 Hz, (b) 10 Hz, (c) 14 Hz, (d) 16 Hz, (e) 18 Hz, and (f) 23 Hz.}
    \label{fig4}
    \end{center}
\end{figure*}

\begin{figure}[ht!]
\begin{centering}
    \includegraphics[width=\columnwidth]{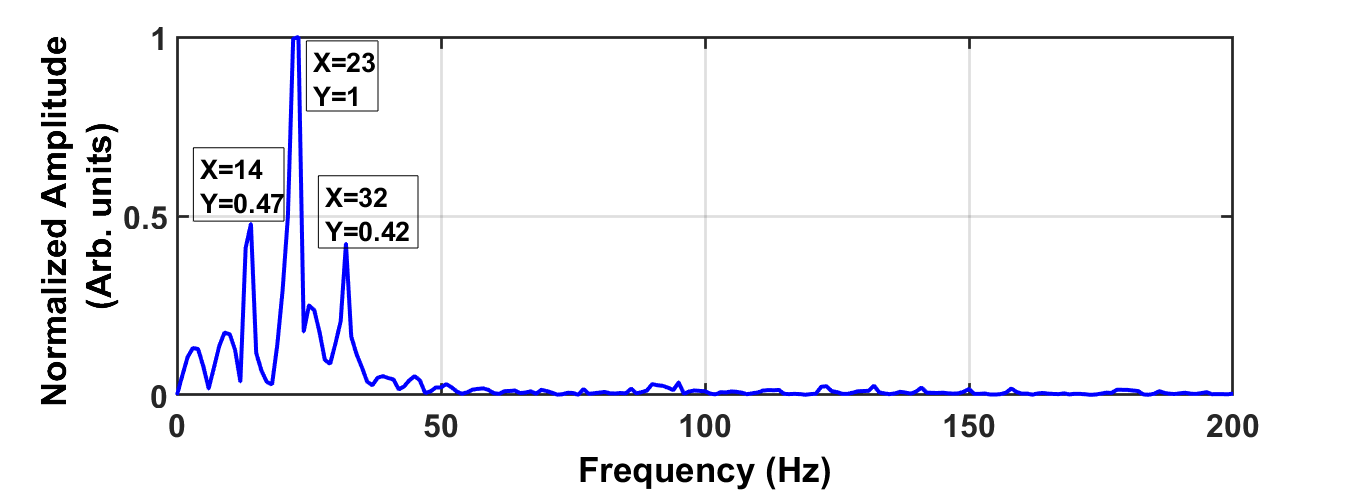}
    \caption{ Frequency spectrum of an unhealthy 20 hp induction motor operating at 23 Hz.}
    \label{fig5}
\end{centering}
\end{figure}

The figures \ref{fig4}(a)-(f) present the FFT spectrum of 20 hp inverter-fed healthy induction motor when it is operated at 8 Hz, 10 Hz, 14 Hz, 18 Hz, and 23 Hz, respectively.It is a healthy induction motor because after applying FFT on the recorded data of 20 hp induction motor, only odd and even components are existed in the harmonics spectrum and no additional sideband frequency component appears in the harmonics spectrum.

Figure \ref{fig5} shows the FFT spectrum of 20 hp inverter-fed unhealthy induction motor when it is operated at 23 Hz. It can be seen that two different sidebands (upper and lower) with equal frequency differences appear around the fundamental frequency component. The upper sideband frequency component is located at 32 Hz with normalized amplitude 0.42 and the lower sideband component is located at 14 Hz with normalized amplitude 0.47. According to equation \ref{equation1}, if any frequency spectrum has equal sidebands with same difference and centred around the fundamental operating frequency then this type of fault is called Broken rotor bar fault. We also have verified this fault by measuring the slip of the 20 hp unhealthy induction motor. Calculated slip is 0.197, lower and upper sideband are calculated by equation \ref{equation1} which is equal to 13.938 Hz and 32.062 Hz, respectively. The total harmonics distortion (THD) of the induction motor is being increased by these sidebands. These sidebands consumed approximately 55\% power of the system. 

\begin{figure}[ht!]
\begin{center}
    \subfigure[]{\includegraphics[width=\columnwidth]{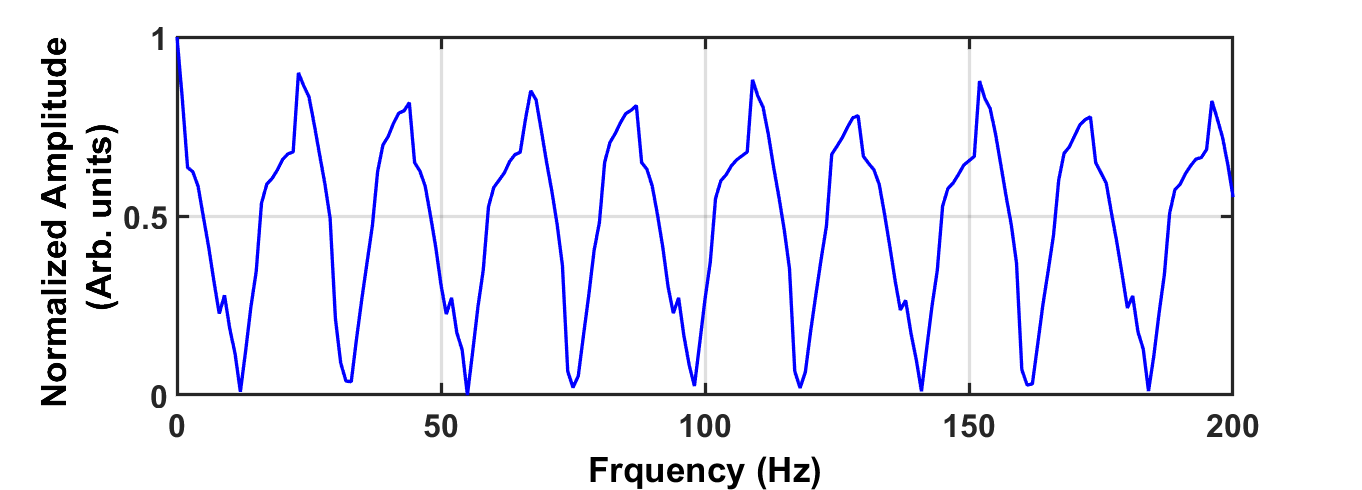}}
    \subfigure[]{\includegraphics[width=\columnwidth]{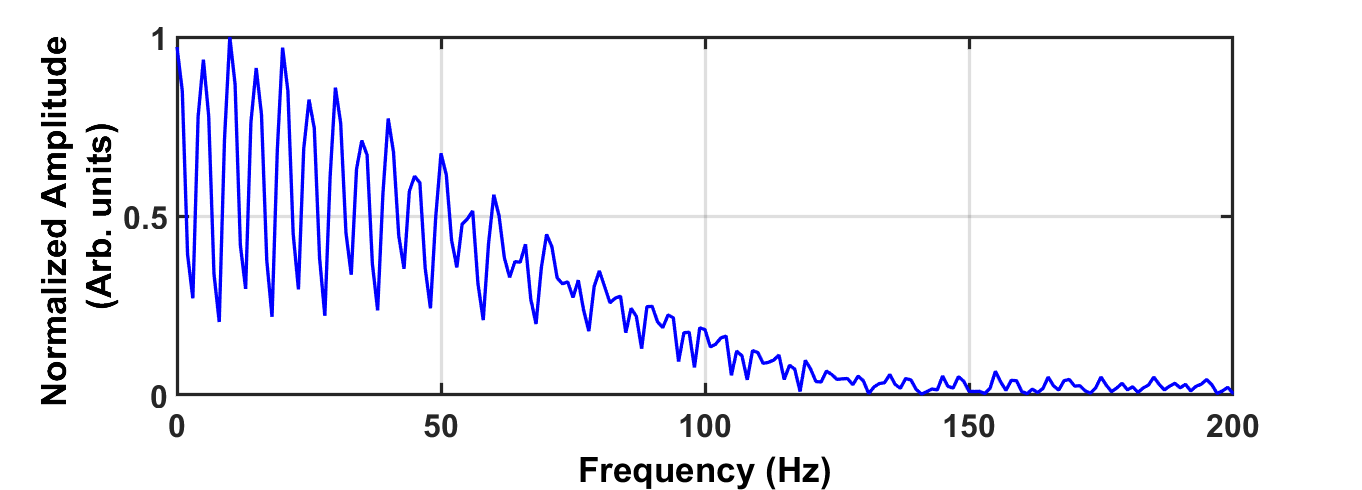}}
    \subfigure[]{\includegraphics[width=\columnwidth]{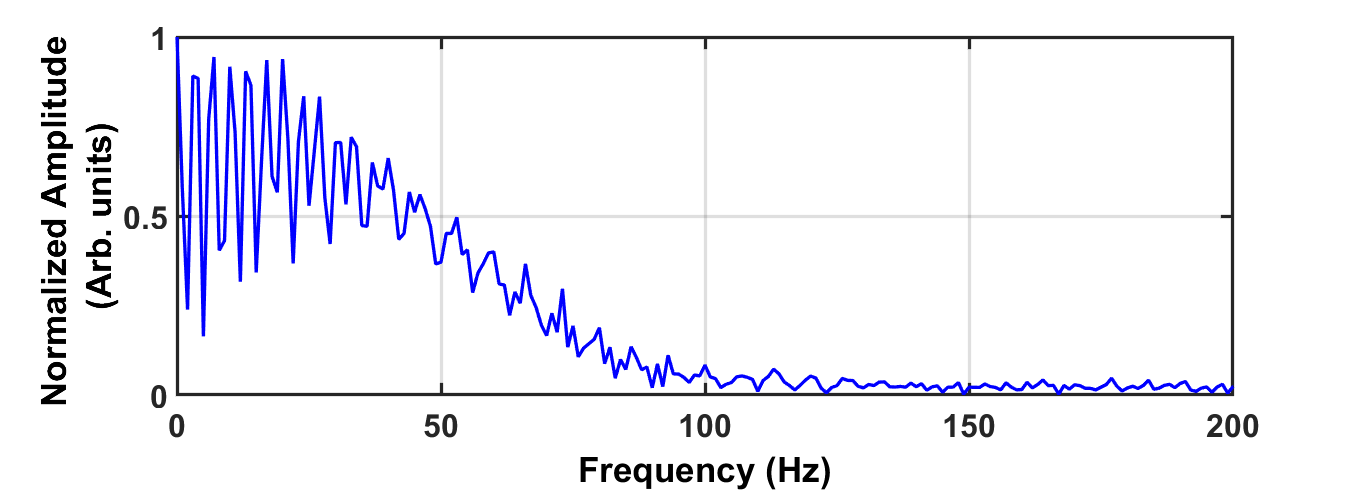}}
    \subfigure[]{\includegraphics[width=\columnwidth]{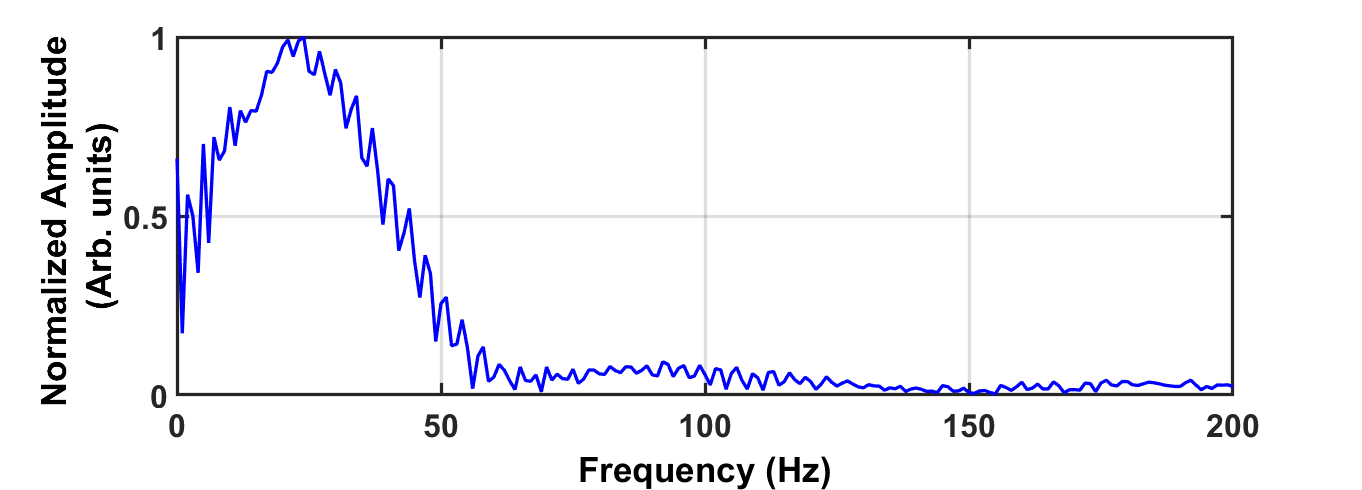}}
    \subfigure[]{\includegraphics[width=\columnwidth]{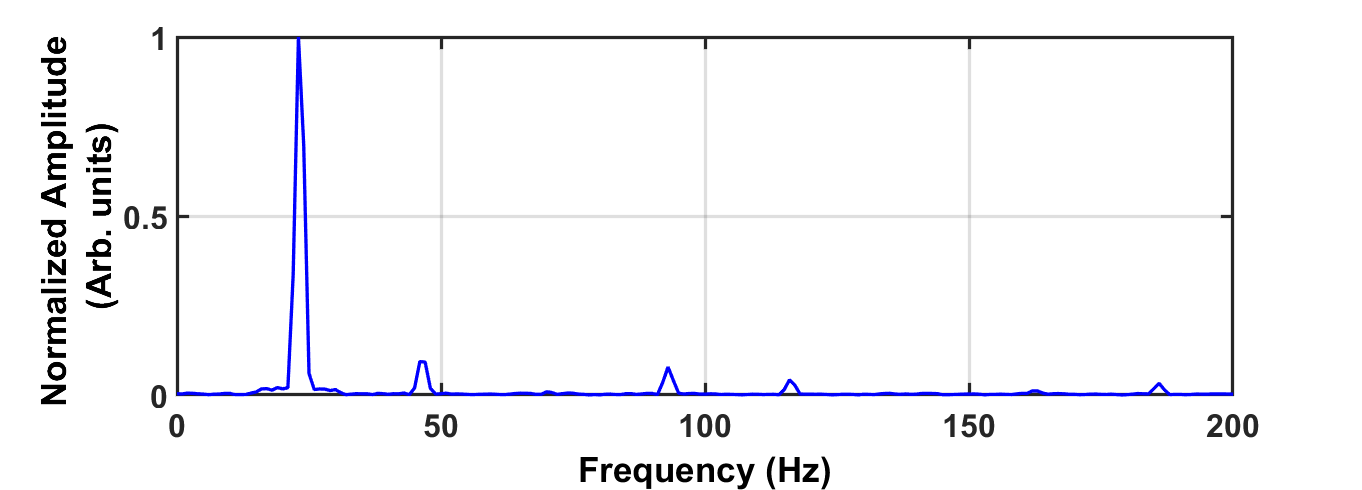}}
    \caption{ Fractional Fourier transform (FrFT) spectrum of healthy 20 hp induction motor operating at 23 Hz with rotational angles set at: (a) 0, (b) 0.85, (c) 0.9, (d) 0.95, and (e) 1.}
    \label{fig6}
    \end{center} 
\end{figure}

The recorded data of healthy 20 hp and 40 hp inverter-fed induction motors has been processed using FrFT by varying the rotational angle from 0 to 1. Figure \ref{fig6}(a) shows the FrFT of 20 hp induction motor with rotational angle 0, whereas figures \ref{fig6}(b)-(e) show the FrFT with varying rotational angle from 0.85 to 1, in a step of 0.05. The rotational angle between 0 to 0.85 is excluded because the FrFT spectrum of both motors is much similar to rotational angle 0. The FrFT spectrum for rotational angle set at 0 is similar to the recorded time-domain signal because, at rotational angle 0, the total sum of kernel function is zero. Whereas for rotational angle set at 0.85 on wards, the FrFT spectrum appears to move from time-domain representation to the frequency domain representation.

\begin{figure}[ht!]
 \begin{center}
    \subfigure[]{\includegraphics[width=\columnwidth]{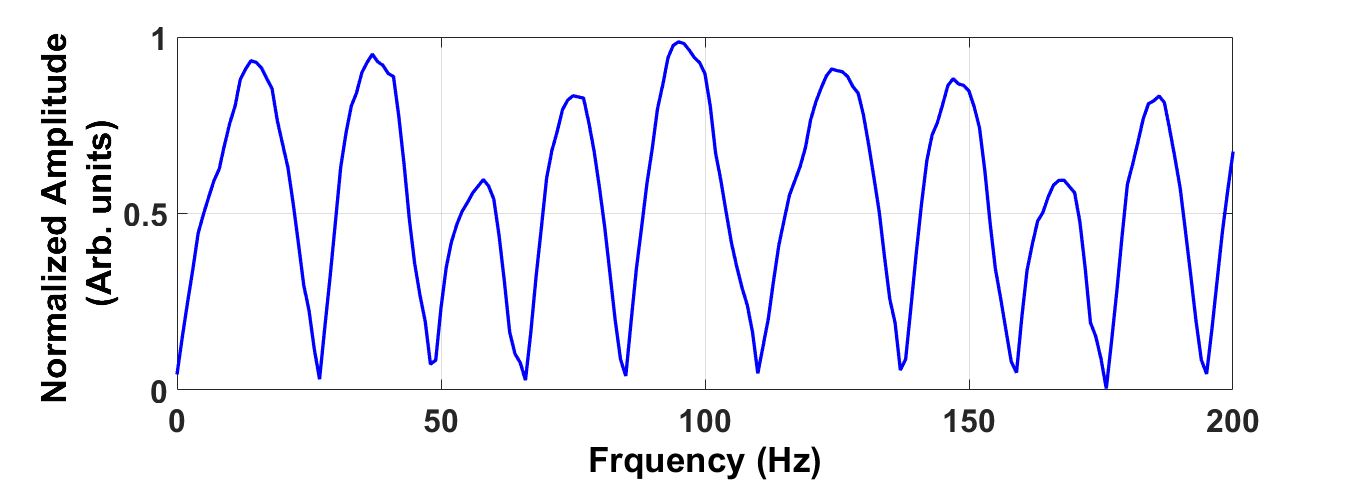}}
    \subfigure[]{\includegraphics[width=\columnwidth]{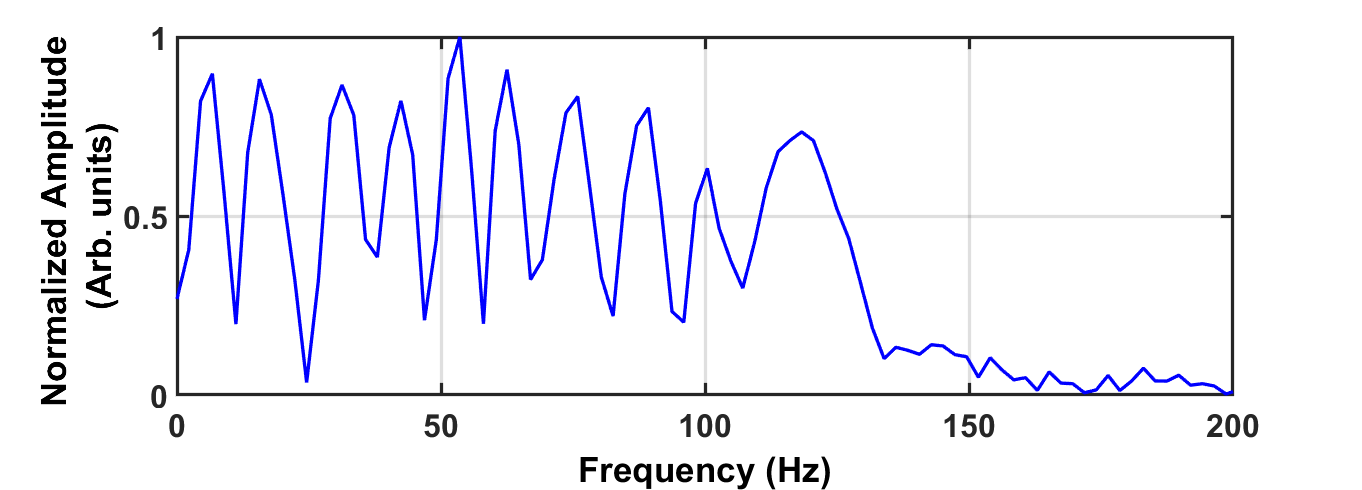}}
    \subfigure[]{\includegraphics[width=\columnwidth]{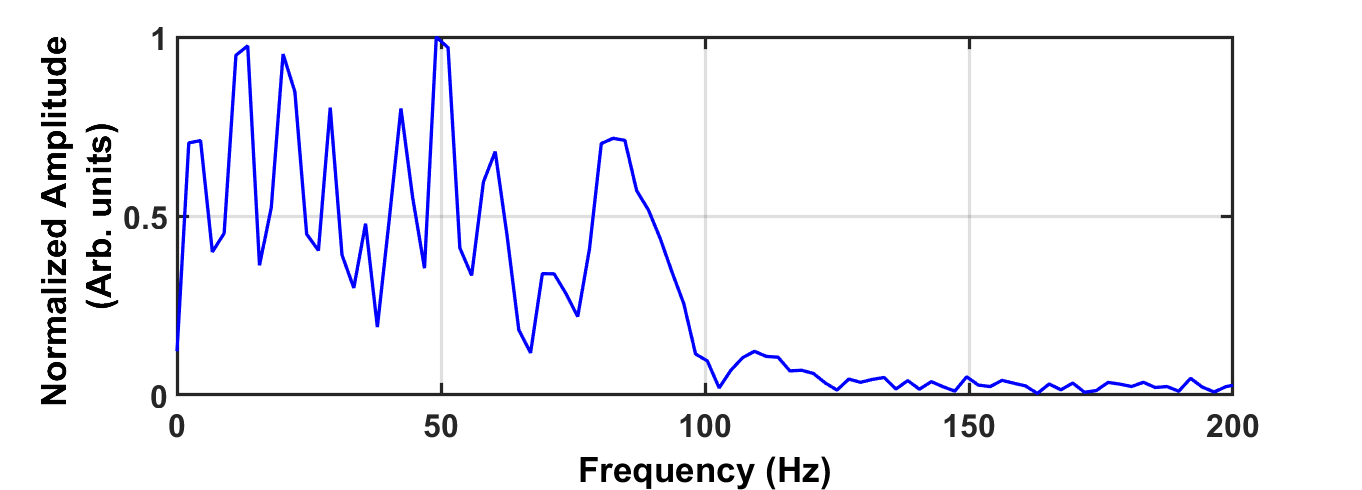}}
    \subfigure[]{\includegraphics[width=\columnwidth]{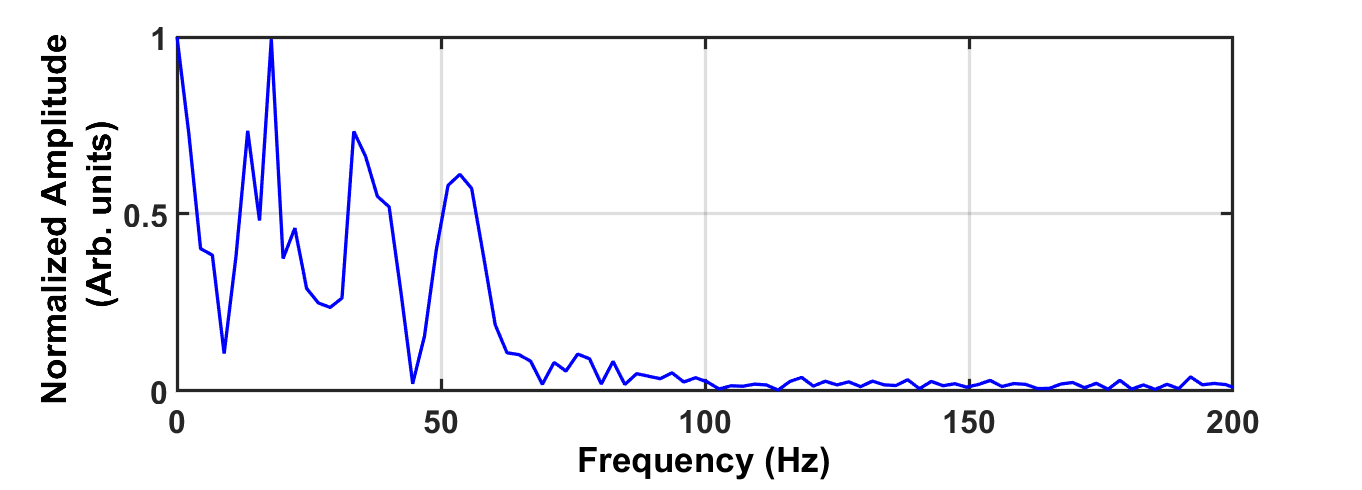}}
    \subfigure[]{\includegraphics[width=\columnwidth]{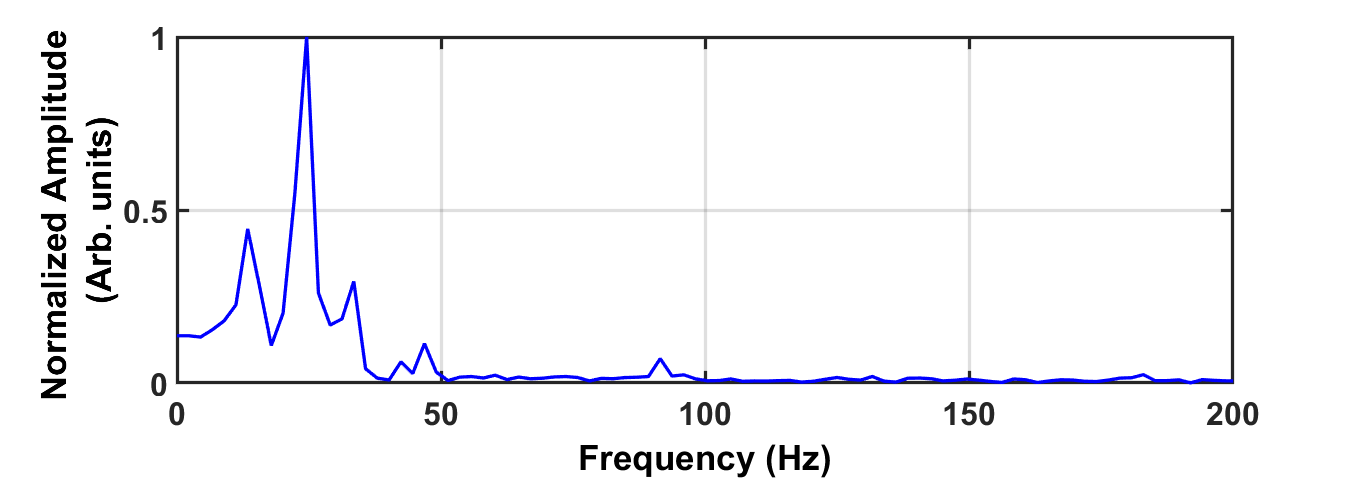}}
    \caption{ Fractional Fourier transform (FrFT) spectrum of unhealthy 20 hp induction motor operating at 23 Hz with rotational angle set at: (a) 0, (b) 0.85, (c) 0.90, (d) 0.95, and (e) 1.}
    \label{fig7}
    \end{center}
\end{figure}

Similarly, the recorded data of unhealthy inverter-fed 20 hp induction motor has been processed using FrFT. Figure \ref{fig7}(a) shows the FrFT spectrum of the unhealthy motor at the rotational angle 0, and figure \ref{fig7}(e) shows the FrFT spectrum at the rotational angle 1. It can be noticed that the FrFT spectrum shown in \ref{fig7}(f) resembles to the FFT spectrum shown in figure \ref{fig5} for the same motor.

\begin{figure}[ht!]
    \subfigure[]{\includegraphics[width=\columnwidth]{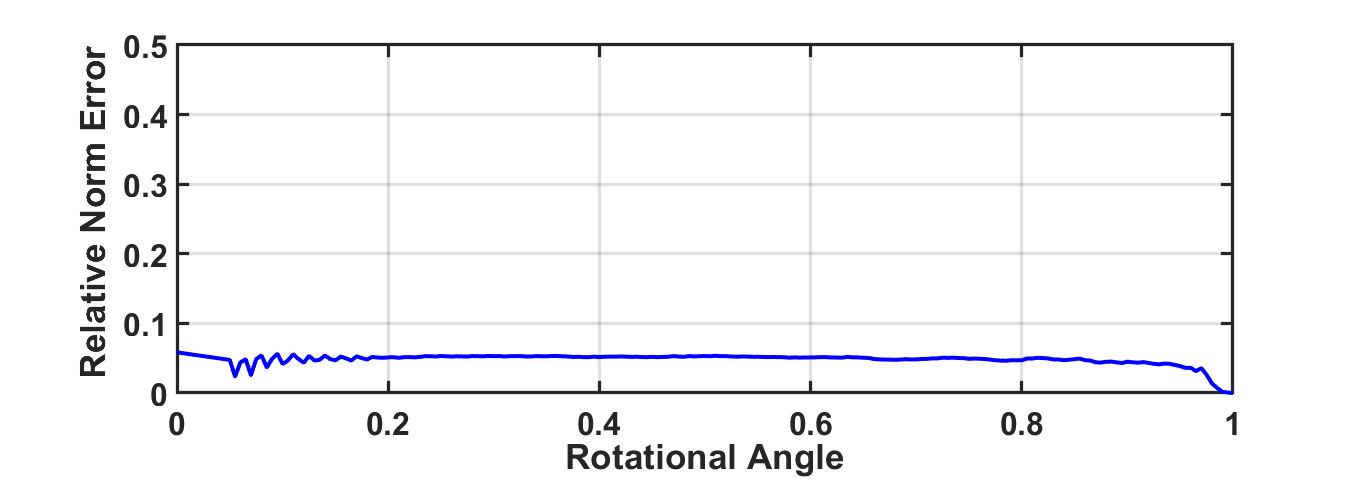}}
    \subfigure[]{\includegraphics[width=\columnwidth]{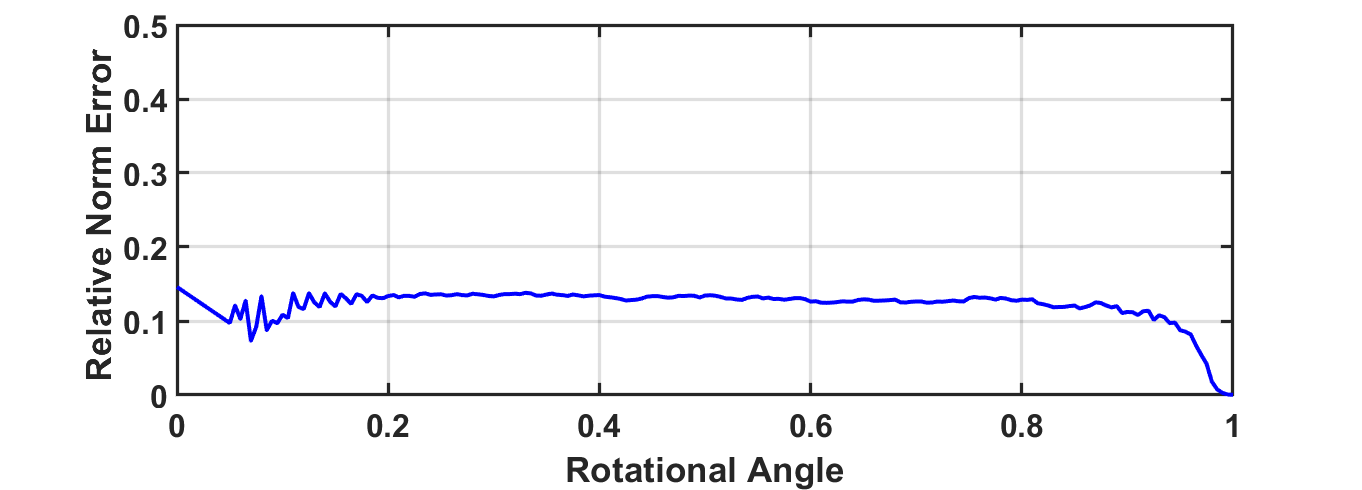}}
    \subfigure[]{\includegraphics[width=\columnwidth]{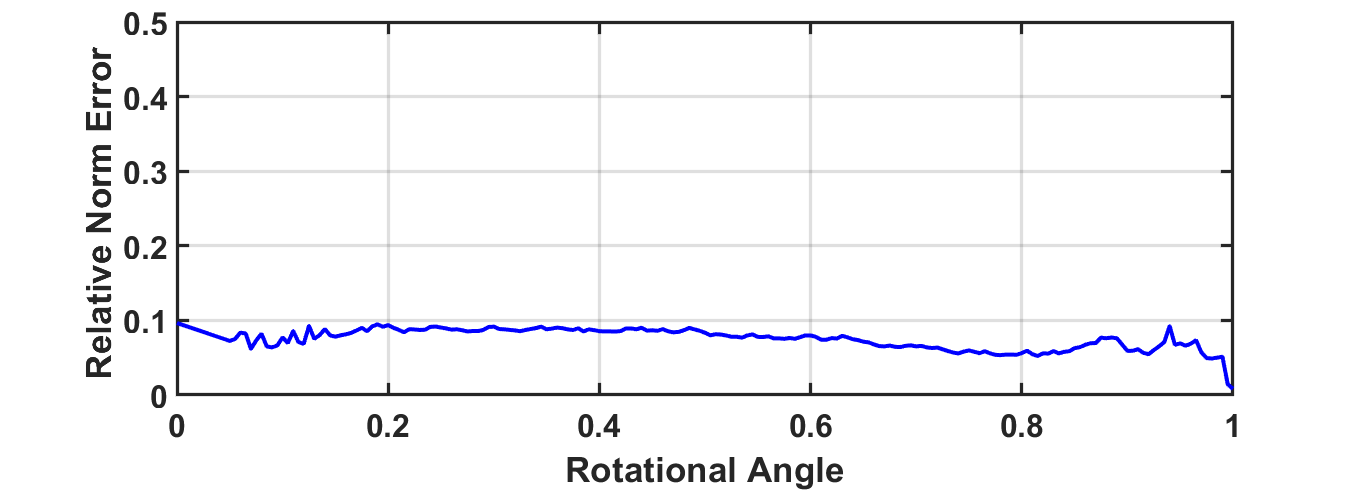}}
    \subfigure[]{\includegraphics[width=\columnwidth]{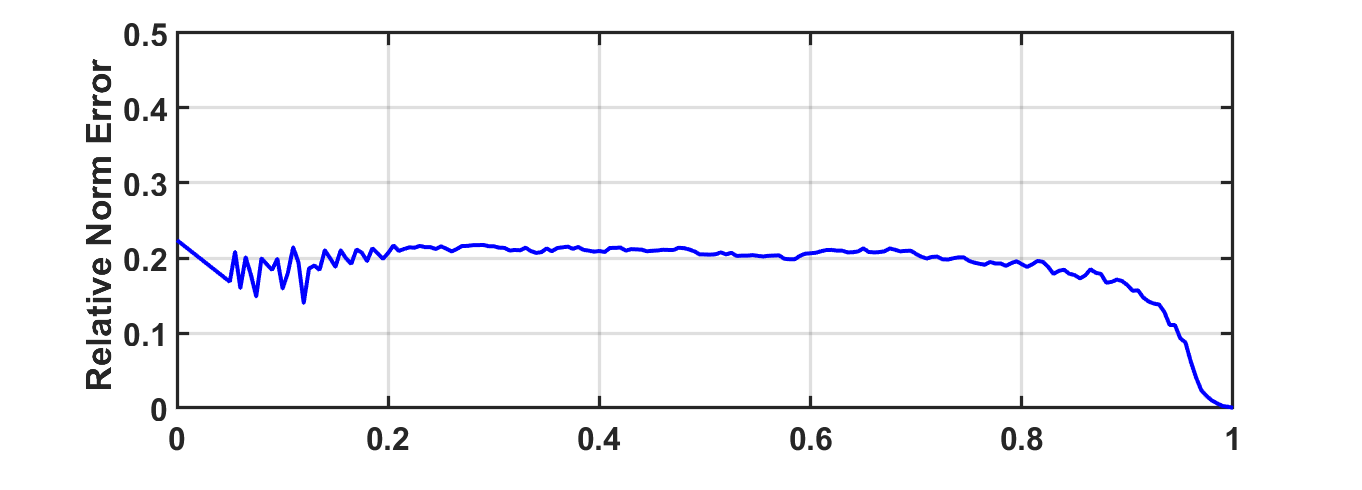}}
    \subfigure[]{\includegraphics[width=\columnwidth]{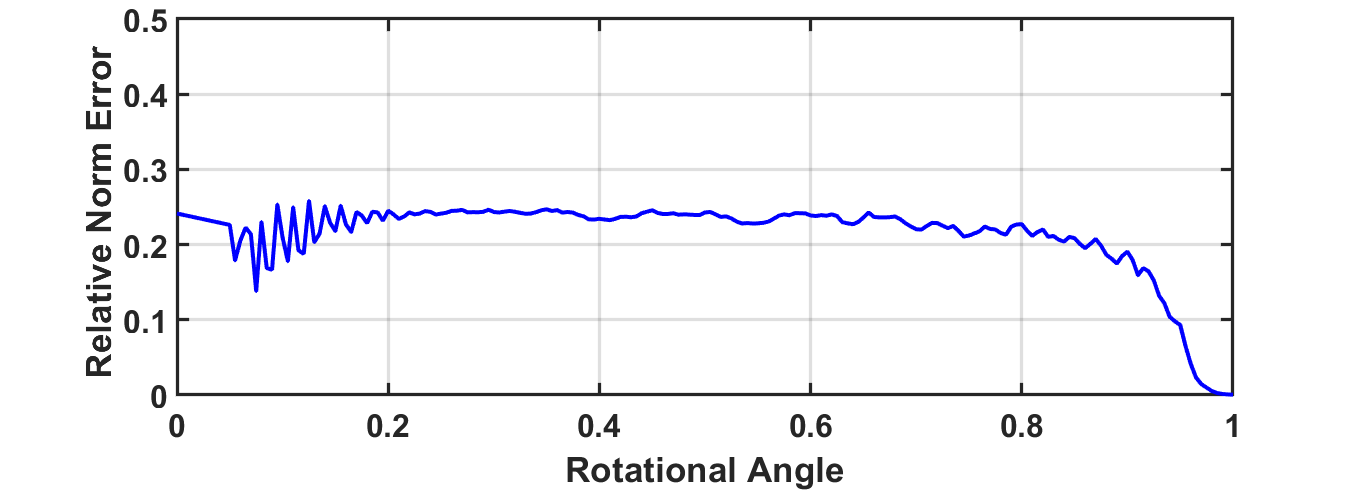}}
    \caption{Relative norm error calculated for the healthy 20 hp induction motor at operating frequencies (a) 8 Hz, (b) 10 Hz, (c) 14 Hz, (d) 18 Hz, and (e) 23 Hz.}
    \label{fig8}
\end{figure}

Subsequent to the FrFT analysis, the relative norm error of the healthy 20 hp and 40 hp induction motors has been calculated using the recorded data. The figure \ref{fig8} shows the relative error norm calculated for the 20 hp healthy induction motor operating at various frequencies \textemdash two equal length samples, obtained from same induction motor have been used to calculate the relative error norm. The starting phase points are same for two samples. In figure \ref{fig8}(a), the operating frequency of the induction motor is 8 Hz. We can see that the relative norm error is nearly zero. The remaining relative norm error plots of 20 hp induction motor are given in figure \ref{fig8}(b)-(e) for varying operating frequencies. We can see that the threshold value of the relative norm error is between 0-0.25 for all the cases.

\begin{figure}[ht!]
    \subfigure[]{\includegraphics[width=\columnwidth]{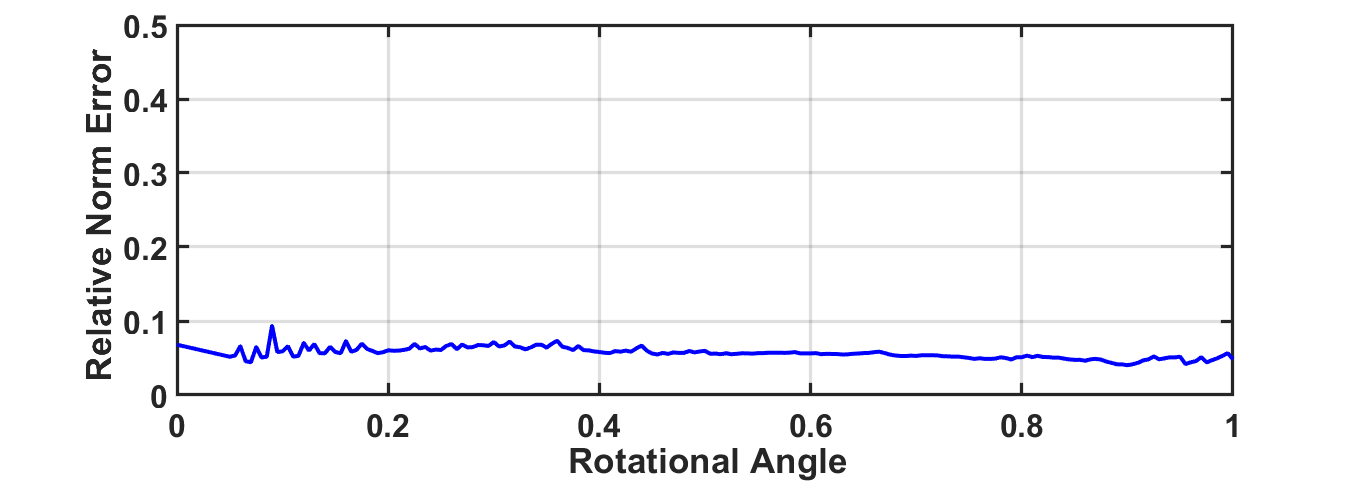}}
    \subfigure[]{\includegraphics[width=\columnwidth]{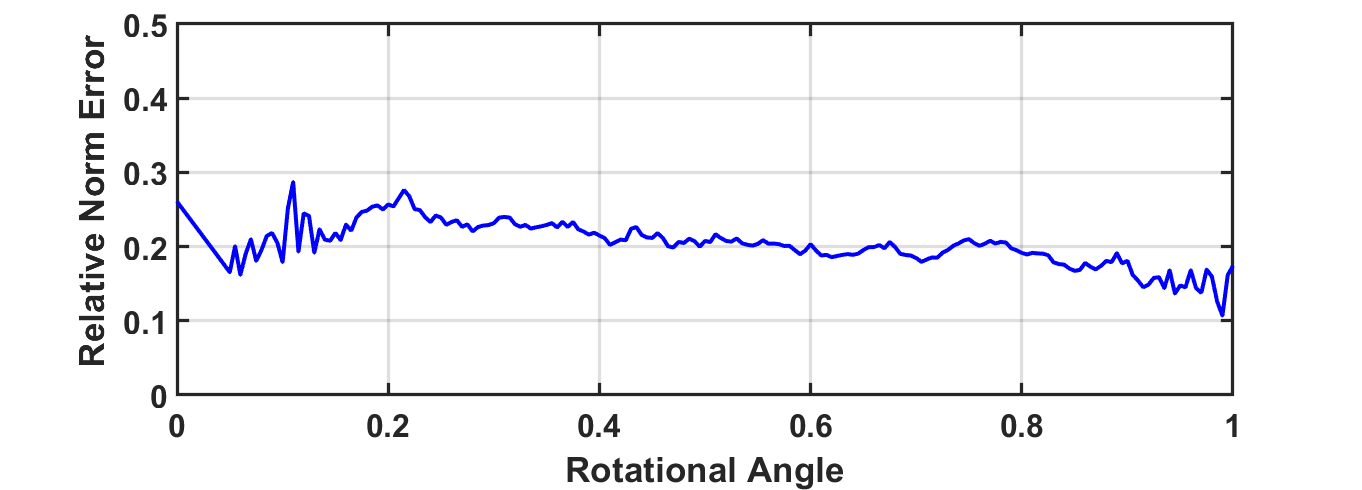}}  
    \subfigure[]{\includegraphics[width=\columnwidth]{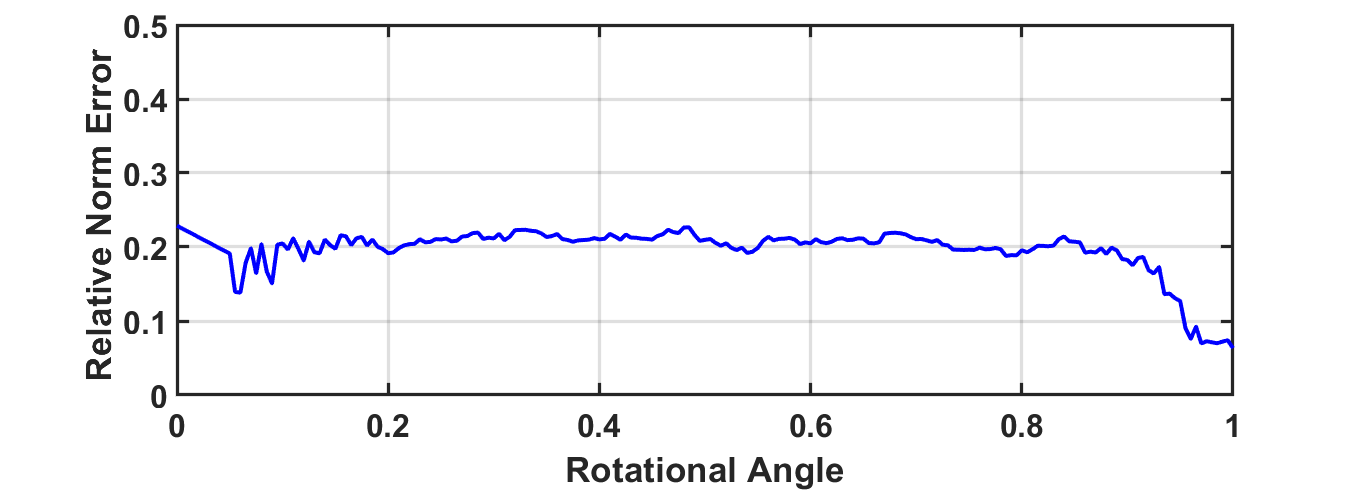}}  
    \subfigure[]{\includegraphics[width=\columnwidth]{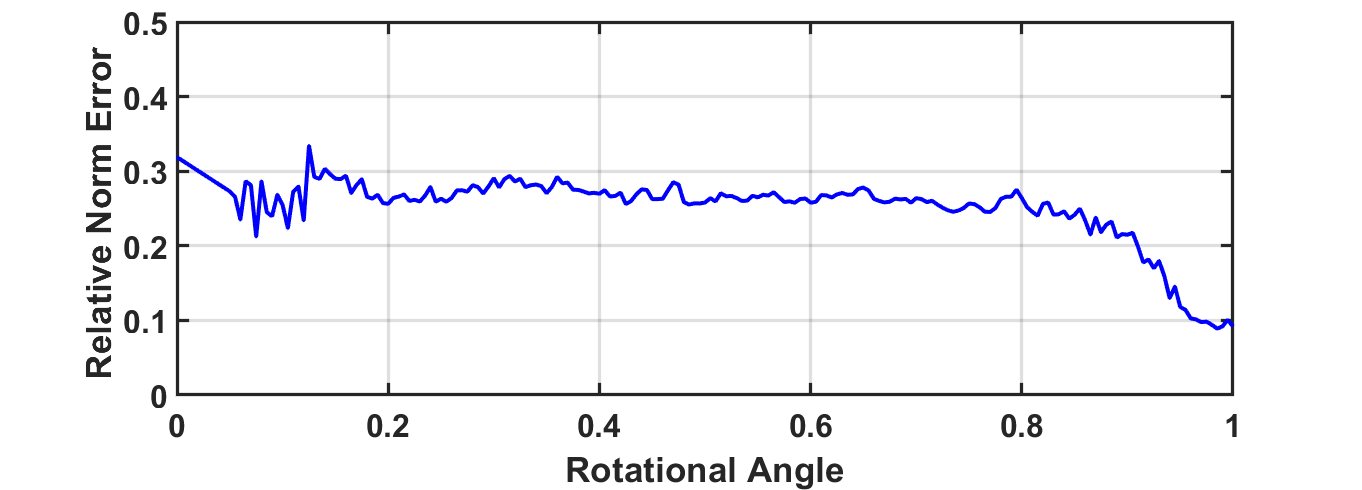}}  
    \subfigure[]{\includegraphics[width=\columnwidth]{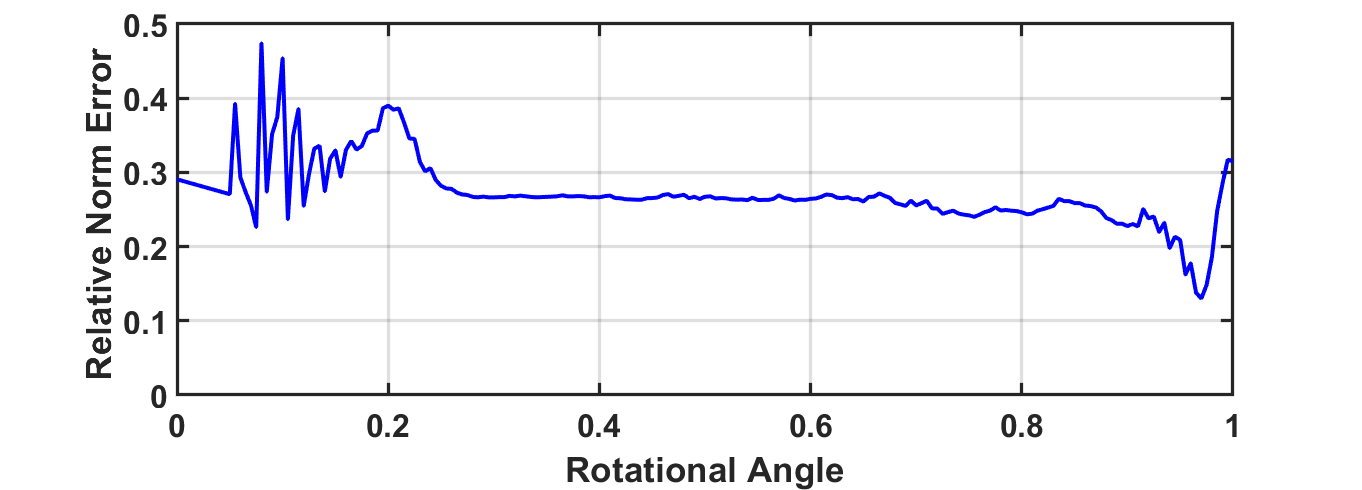}}  
   \caption{Relative norm error calculated for the healthy 40 hp induction motor operating at frequencies (a) 8 Hz, (b) 10 Hz, (c) 14 Hz, (d) 18 Hz, and (e) 23 Hz.}
    \label{fig9}
\end{figure}

Similarly, the calculations for the relative norm error have been performed for healthy 40 hp induction motor, and presented in figure \ref{fig9}. Figure \ref{fig9}(e) shows that the maximum relative error norm value when the induction motor is operated at 23Hz, whereas, figure \ref{fig9}(A) shows the lowest relative error norm, nearly zero, when the motor is operating at 8 Hz. The  threshold value of the relative norm error is between 0-0.3 for the remaining cases as shown in figures \ref{fig9}(b)-(d).

\begin{figure}[ht!]
    \subfigure[]{\includegraphics[width=\columnwidth]{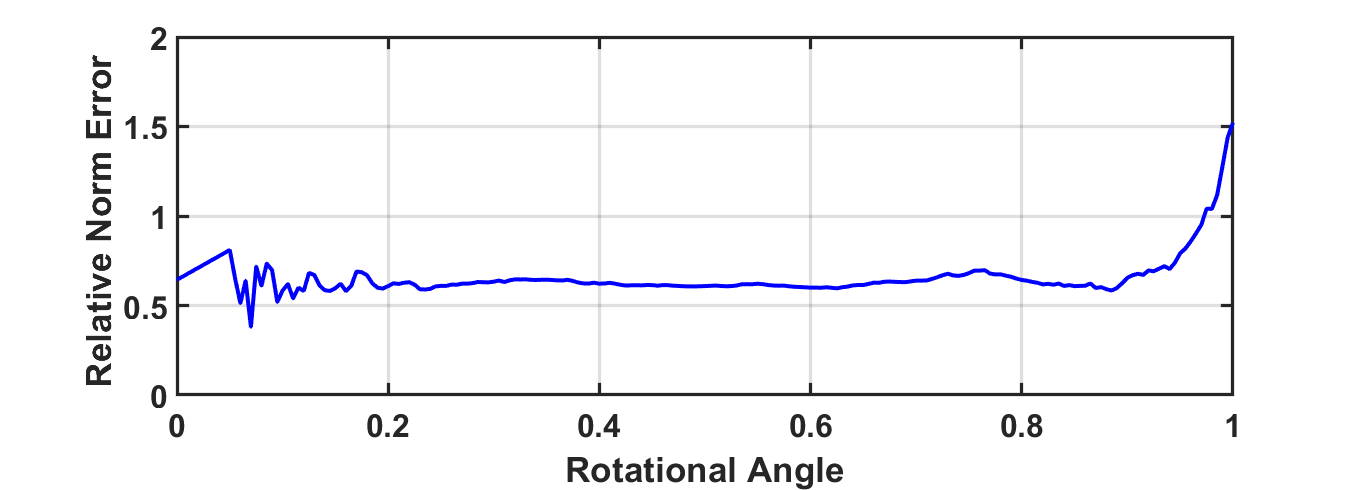}}
    \subfigure[]{\includegraphics[width=\columnwidth]{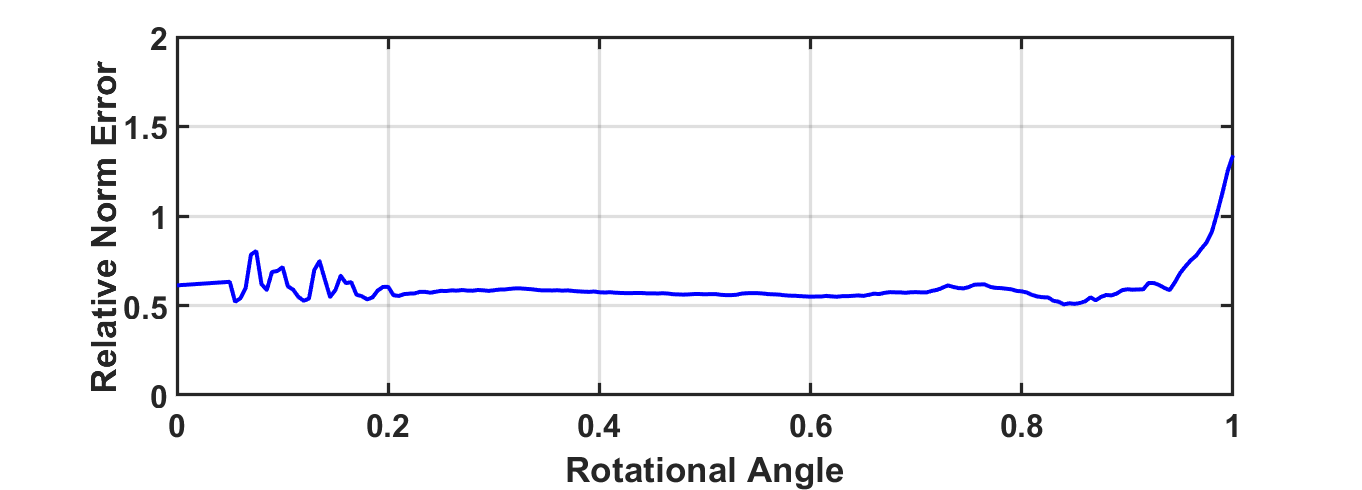}}  
    \subfigure[]{\includegraphics[width=\columnwidth]{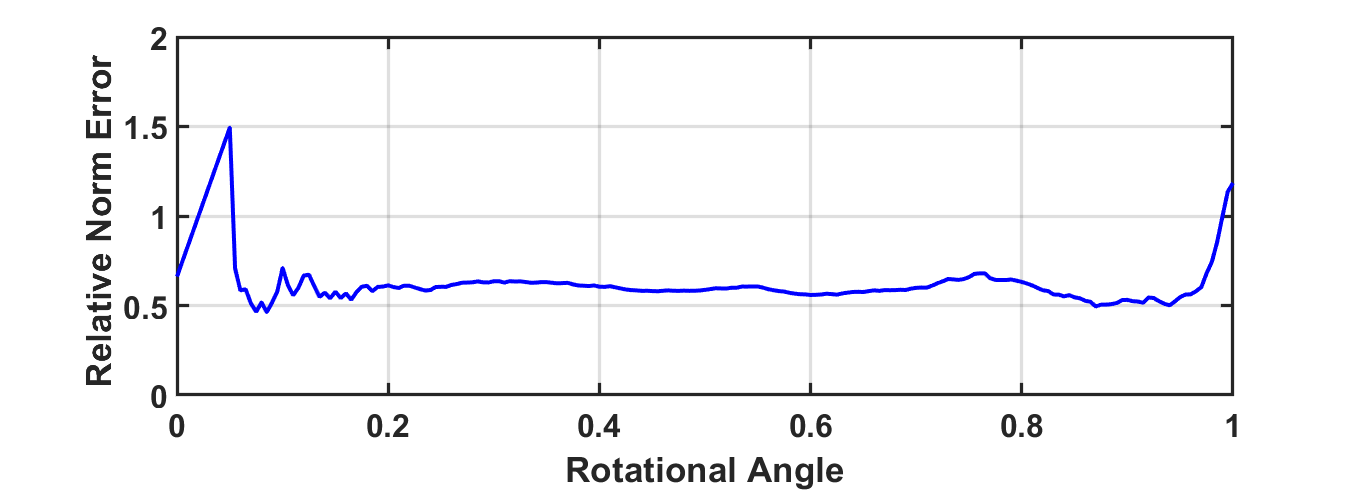}}  
    \subfigure[]{\includegraphics[width=\columnwidth]{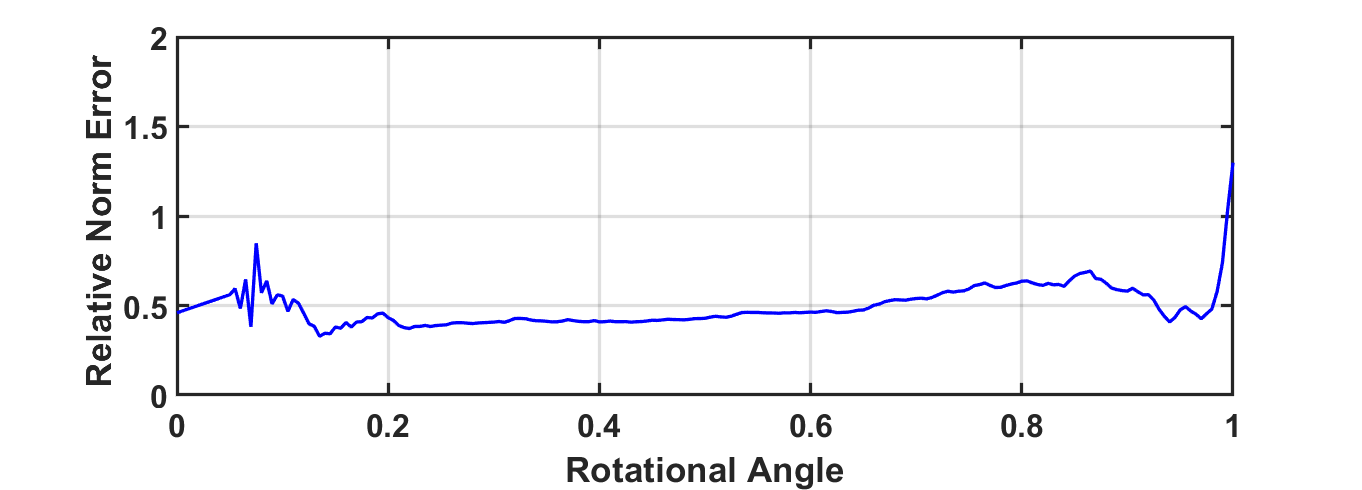}}  
    \subfigure[]{\includegraphics[width=\columnwidth]{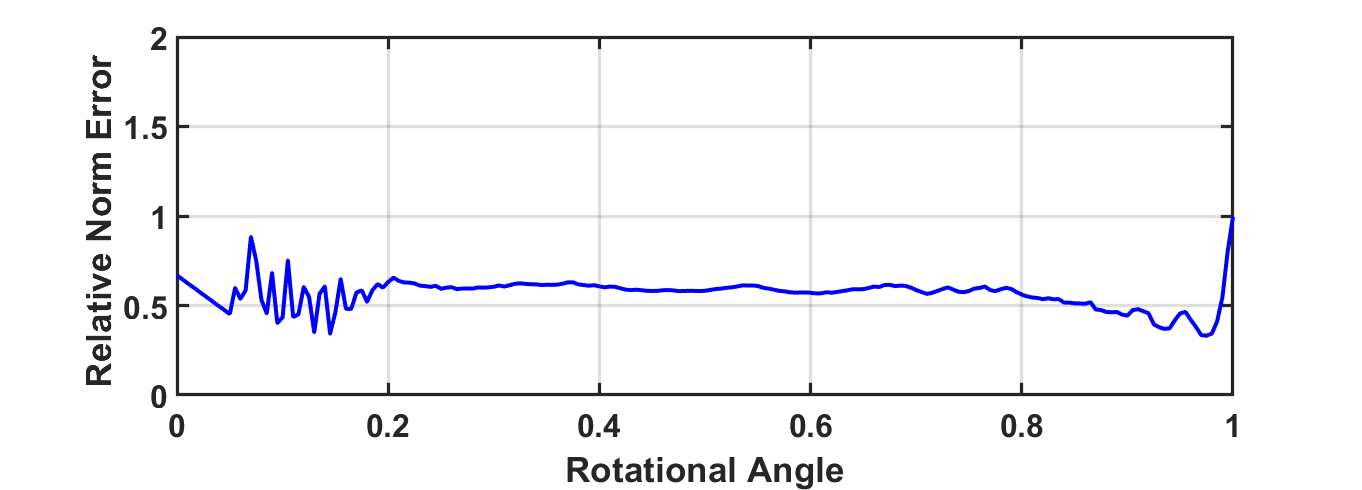}}  
   \caption{Relative norm error between unhealthy 20 hp induction motor operated at fixed frequency of 23 Hz, and healthy 20 hp motor operated at frequencies: (a) 8 Hz, (b) 10 Hz, (c) 14 Hz, (d) 18 Hz, and (e) 23 Hz.}
    \label{fig10}
\end{figure}

In addition, we have calculated the relative error norm between healthy and unhealthy induction motors. In this experiment, we picked one reference sample of unhealthy 20 hp faulty induction motor operated at 23 Hz and different samples of 20 hp healthy induction motor operated at varying frequencies. Figure \ref{fig10}(a) shows the relative norm error when the healthy induction motor is operated at 8 Hz, whereas operating frequency of unhealthy motor is fixed at 23 Hz. We can see that the relative norm error is in the range of 0.45-0.75 for the rotational angle between 0 to 0.2, whereas the relative norm error becomes constant at a value larger than 0.5 for the rotational angle 0.2-0.85. The error between both motors is maximum at rotational angle 1. Similar pattern can be seen for the variation in relative error norm, calculated at different rotational angles, when the unhealthy motor is operated at fixed frequency of 23 Hz and healthy induction motor is operated at varying frequencies, as shown in figures \ref{fig10}(b)-(e). The average threshold value of relative norm error between healthy and unhealthy induction motor is equal to $\sim$0.5.

\begin{figure}[ht!]
\begin{centering}
    \includegraphics[width=\columnwidth]{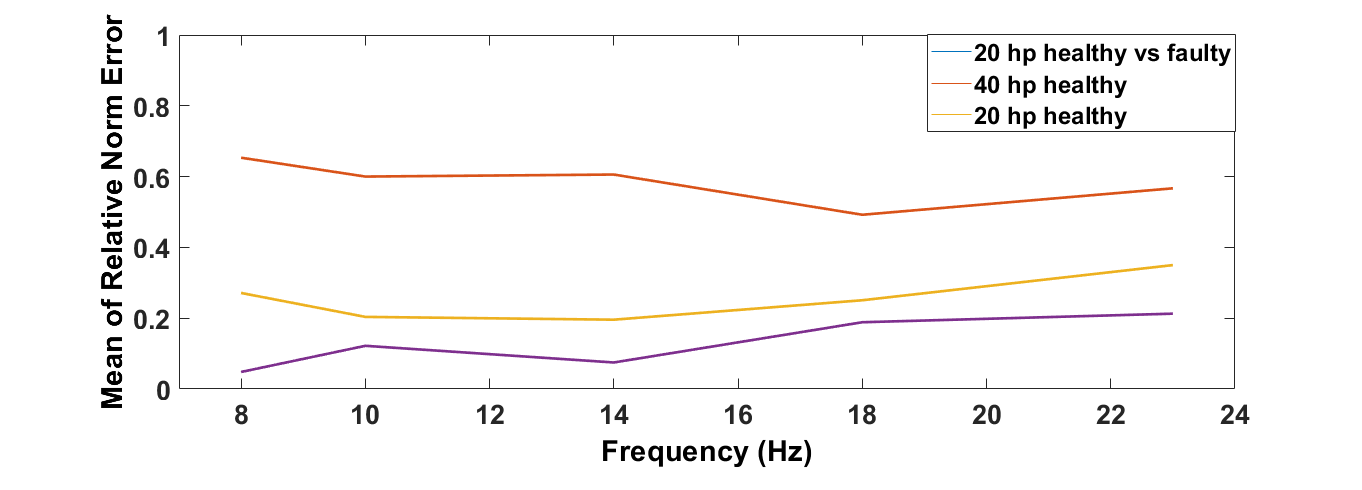}
    \caption{ Mean of relative norm error plot of healthy and unhealthy induction motor at different frequencies. Orange and blue line presents the mean of relative norm error between sampled data of 20 hp and 40 hp healthy induction motor. Red line shows the mean of relative norm error between 20 hp healthy and 20 hp faulty induction motor.}
    \label{fig11}
\end{centering}
\end{figure}

The mean of relative norm error plot concerning healthy and unhealthy induction motors at different operating frequencies is shown in figure \ref{fig11}. It can be seen the mean threshold value of the relative norm error between sampled data of 20 hp and 40 hp healthy induction motor is less than 0.3, and the threshold value between 20 hp healthy and faulty induction motor is greater than 0.5.
 
\section{conclusion}
This paper presents a method for fault diagnosis of a steady-state current signal of inverter-fed induction motor using FFT and FrFT techniques. The developed system has been able to classify the healthy and unhealthy induction motor by calculating the relative norm error and mean threshold values. The experimental finding indicates that the relative norm error threshold value of the healthy induction motor is less than that of an unhealthy induction motor. The threshold values of healthy induction motors have been documented to be less than 0.3, and threshold values between healthy and unhealthy induction motors are greater than 0.5 at various operating frequencies. The developed approach will act as a simple operator-assisted instrument for real-time determination of induction motor faults. Furthermore, the suggested approach can also be extended to other related problems with motor diagnostics.
\label{concl}

\bibliography{References}
\bibliographystyle{IEEEtran}
\EOD
\end{document}